\documentclass{aa}

 \usepackage{ gensymb }
 \usepackage{graphicx}
 \usepackage{natbib,twoopt}
 \usepackage{amsmath}
 \usepackage{ textcomp }
 \usepackage[breaklinks=true]{hyperref} 
 \bibpunct{(}{)}{;}{a}{}{,}    
 \newcommandtwoopt{\citeads}[3][][]{\href{http://adsabs.harvard.edu/abs/#3}%
                                        {\citealp[#1][#2]{#3}}}
 \newcommandtwoopt{\citepads}[3][][]{\href{http://adsabs.harvard.edu/abs/#3}%
                                        {\citep[#1][#2]{#3}}}
 \newcommandtwoopt{\citetads}[3][][]{\href{http://adsabs.harvard.edu/abs/#3}%
                                        {\citet[#1][#2]{#3}}}
 \newcommandtwoopt{\citeyearads}[3][][]%
   {\href{http://adsabs.harvard.edu/abs/#3}{\citeyear[#1][#2]{#3}}}

\begin{document}
\titlerunning{Density models}
\authorrunning{Zucca et al.}
\title{The formation heights of coronal shocks from 2D density and Alfv\'en speed maps}
\author{Pietro Zucca, Eoin P. Carley, D. Shaun Bloomfield and Peter T. Gallagher }
\institute{Astrophysics Research Group, School of Physics, Trinity College Dublin, Dublin 2, Ireland.}

\abstract{Super-Alfv\'enic shocks associated with coronal mass ejections (CMEs) can produce radio emission known as Type II bursts. In the absence of direct imaging, accurate estimates of coronal electron densities, magnetic field strengths, and Alfv\'en speeds are required to calculate the kinematics of shocks. To date, 1D radial models have been used, but these are not appropriate for shocks propagating in non--radial directions.} {Here, we study a coronal shock wave associated with a CME and Type II radio burst using 2D electron density and Alfv\'en speed maps to determine the locations that shocks are excited as the CME expands through the corona.}{Coronal density maps were obtained from emission measures derived from the Atmospheric Imaging Assembly (AIA) on board the \emph{Solar Dynamic Observatory} ($SDO$) and polarized brightness measurements from the Large Angle and Spectrometric Coronagraph (LASCO) on board the \emph{Solar and Heliospheric Observatory} ($SOHO$). Alfv\'en speed maps were calculated using these density maps and magnetic field extrapolations from the Helioseismic and Magnetic Imager ($SDO$/HMI). The computed density and Alfv\'en speed maps were then used to calculate the shock kinematics in non-radial directions.}{Using the kinematics of the Type II burst and associated shock, we find our observations to be consistent with the formation of a shock located at the CME flanks where the Alfv\'en speed has a local minimum.}{The 1D density models are not appropriate for shocks that propagate non--radially along the flanks of a CME. Rather, the 2D density, magnetic field and Alfv\'en speed maps described here give a more accurate method for determining the fundamental properties of shocks and their relation to CMEs.}
\maketitle

\section{Introduction}
Solar flares and coronal mass ejections (CMEs) are energetic manifestations of the restructuring coronal magnetic fields. As a CME travels through the corona, its velocity can become sufficiently larger than the background coronal Alfv\'en speed, causing a shock wave to form along its leading edge and/or flanks \citep{Cho2007}. It is within these shocks that electrons can be accelerated to near-relativistic energies to produce Type II radio signatures in low-frequency dynamic spectra \citep[see,][]{Carley2013}. Although CMEs and Type II radio bursts have been studied for many decades \citep[see,][]{Pick2006}, there remains unanswered questions in relation to where CME shocks are formed and how these phenomena are associated with the generation of Type II bursts.

It has long been suggested that Type II radio bursts are signatures of coronal shocks \citep{Wild1950,Uchida1960}. They appear as features slowly drifting toward lower frequencies at decimetric to kilometric wavelengths in dynamic radio spectra. This drift is the result of plasma emission generated by a super-Alfv\'enic shock traveling upwards in the corona \citep{Cane1981}, where density decreases with height. When direct low-frequency imaging is not available, a key problem is therefore the accurate calculation of the coronal density and Alfv\'en speed distributions with height; This is to relate the Type II emission frequency to its height and to investigate the direction of the shock propagation.
\begin{figure}[h]
\begin{center}
\includegraphics[trim=1cm 1cm 1cm 1cm,clip=true,width=9cm,angle=0]{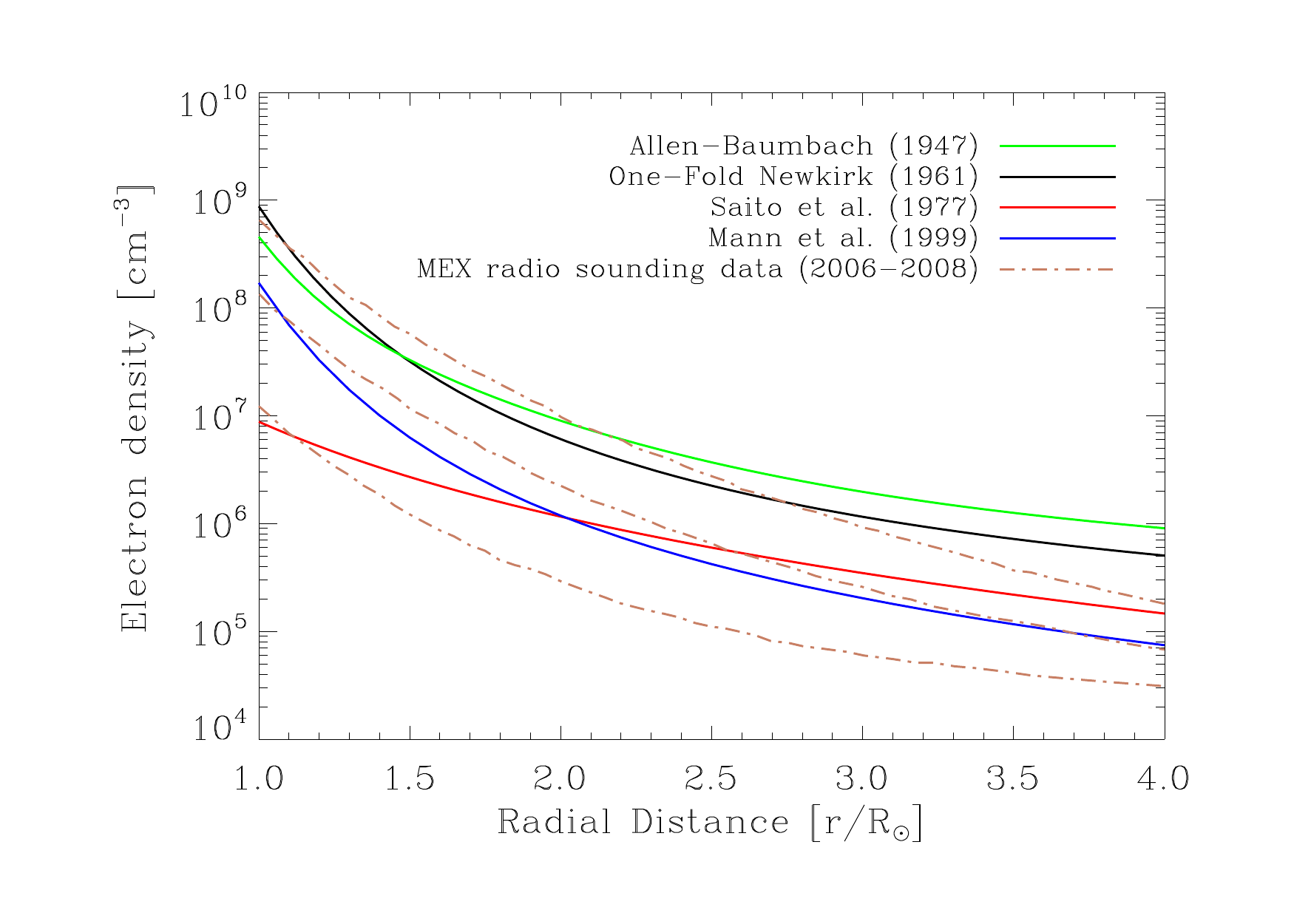}
\caption{A variety of commonly used electron density models, which can give very different height estimates. For example, a density of $10^7$ cm$^{-3}$ can be located at $\sim$1.4~$R_{\odot}$ or $\sim$2.0~$R_{\odot}$, depending on the density model. Plotted for comparison, the set of density measurements obtained with Mars Express (MEX) between the years 2006--2008 \citep{Verma2013}.}
\label{Fig_1_models}
\end{center}
\end{figure}
The plasma frequency is related to the density of the emitting plasma by $f_\mathrm{p}= C\sqrt{n_\mathrm{e}} $, where $C=8980$ Hz cm$^{3/2}$ is a constant. To derive the shock kinematics, electron density models are normally employed to relate the plasma density to its coronal height and velocity. Specifically, the shock radial velocity is related to the plasma frequency drift rate, $\mathrm{d} f_\mathrm{p}/\mathrm{d} t$,  and the electron density model, $ n_\mathrm{e}(r)$,  by,
\begin{equation}
v= \frac{2\sqrt{n_\mathrm{e}}}{C} \left( \frac{\mathrm{d} n_\mathrm{e}}{\mathrm{d} r} \right)^{-1} \frac{\mathrm{d} f_\mathrm{p}}{\mathrm{d} t} ,
\end{equation}
\begin{figure}
\centering
\includegraphics[trim=0cm 0cm 0cm 0cm, clip=true, width=9cm, angle=0]{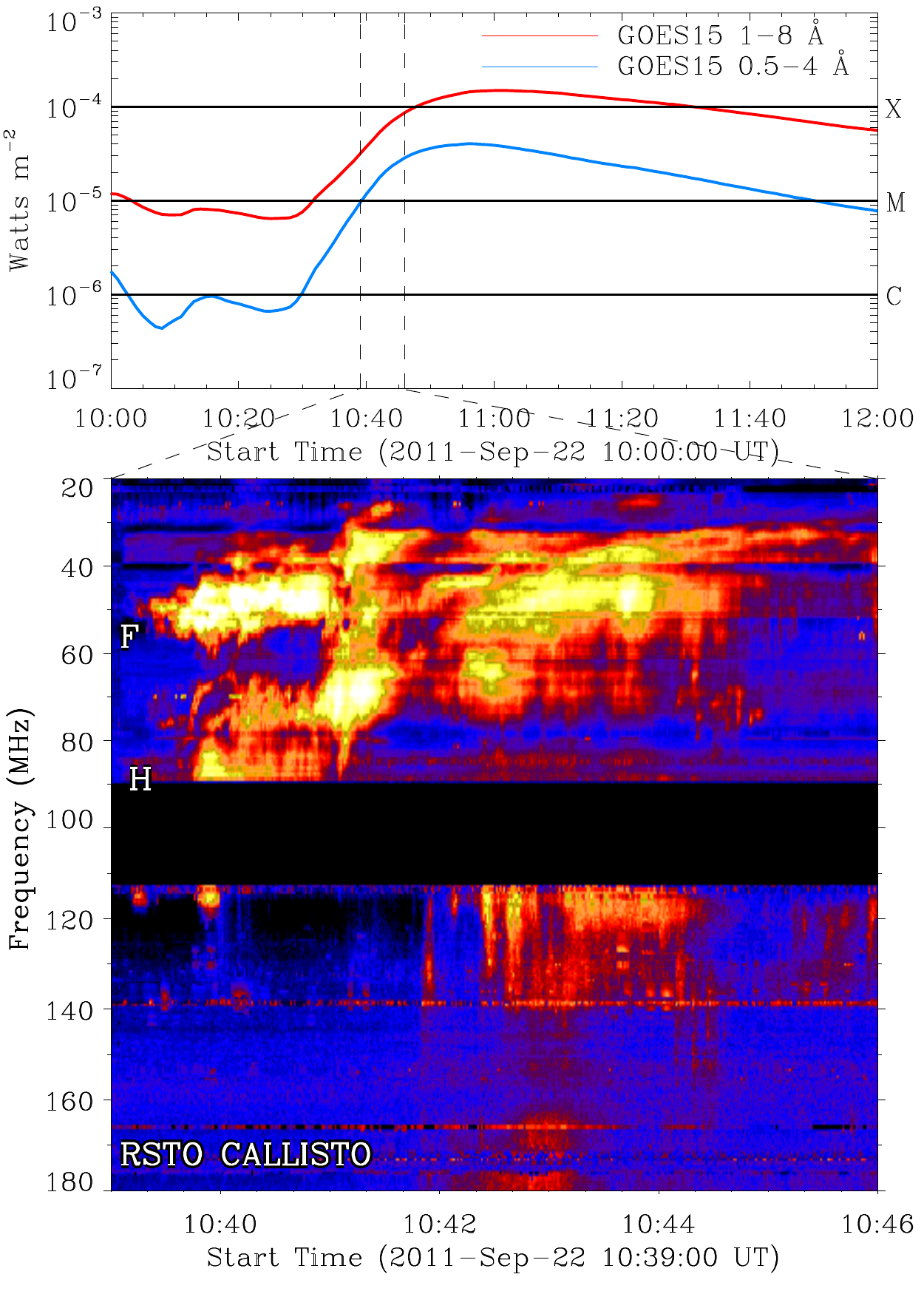}
\caption{\emph{GOES--15} soft X--ray light curve showing an X1.4 flare starting at 10:29:00~UT on 2011 Septemeber 22 (top) and the associated RSTO dynamic spectrum showing a Type II radio burst starting at 10:39:06~UT (bottom). This burst shows fundamental (F) and harmonic (H) emission.}
\label{RSTO}
\end{figure}
where $v$ is the shock velocity, $n_\mathrm{e}$ is the coronal plasma electron density and $r$ is the heliocentric radial distance. Different coronal density models can therefore lead to differing kinematics. Several density models have been used to calculate the Type II shock position, such as the \citet{Newkirk1961} model derived from the barometric height behavior of a gravitationally stratified corona, the \citet{Saito1977} model obtained from measurements of coronal polarized brightness ($pB$), and the \citet{Mann1999} model, derived from solutions of magnetohydrodynamic equations.  

The use of an arbitrary radial density model can lead to an inaccurate calculation of shock heights and hence velocities. This is due to the significant difference in the shock height derived from different density models. An example is shown in Fig.~\ref{Fig_1_models}, where an electron density of $10^7$~cm$^{-3}$, occurs at a height of 1.4~$R_{\odot}$ for the \citet{Mann1999} model,  while occurs at a height of 2.0~$R_{\odot}$ for the Allen-Baumbach model \citep{Allen1947}. Another reason for an inaccurate shock height and velocity calculation is the time variability of the coronal density distribution \citep{Parenti2000,Bemporad2003}, which is not taken into account with typically employed density models. Finally, if a Type II radio source propagates non-radially, speeds derived from 1D radial density models underestimate the true shock speed. 

Knowledge of the coronal density in the 2D plane is therefore crucial for determining accurate shock kinematics, while the knowledge of the 2D Alfv\'en speed is important to determine when the propagating shock reaches a super-Alfv\'enic speed. A 2D analytic model of the Alfv\'en speed was presented by \citet{Warmuth2005}. However, due to the temporal variability of the density and magnetic field in the corona, it is important to use observational data specific to the radio emission time rather than a generic analytic model for the Alfv\'en speed. 
\begin{figure*}
\begin{center}
\includegraphics[trim=6.1cm 2.5cm 6.5cm 2.9cm,clip=true,width=6.7cm,angle=90]{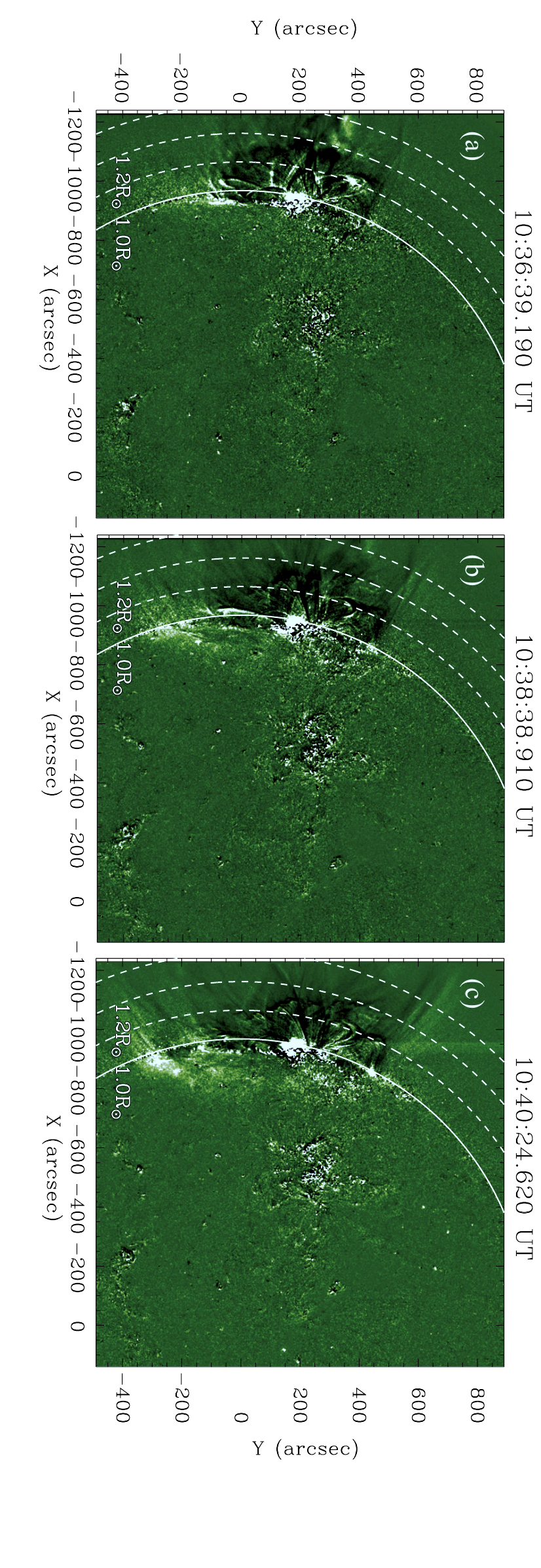}
\caption{{\it SDO}/AIA running--difference images of the erupting CME at 10:36:39--10:40:24~UT in the 211~\AA\ passband. The CME leading edge is outside the field--of--view of $SDO$ in panel (a), while the propagation of the coronal bright front is visible at the location of the CME flanks at 10:38:38~UT panel (b) and at 10:40:24~UT panel (c).}
\label{SDO_cme}
\end{center}
\end{figure*}

\begin{figure*}
\begin{center}
\includegraphics[trim=6cm 3.8cm 5.9cm 4cm,clip=true,width=7.9cm,angle=90]{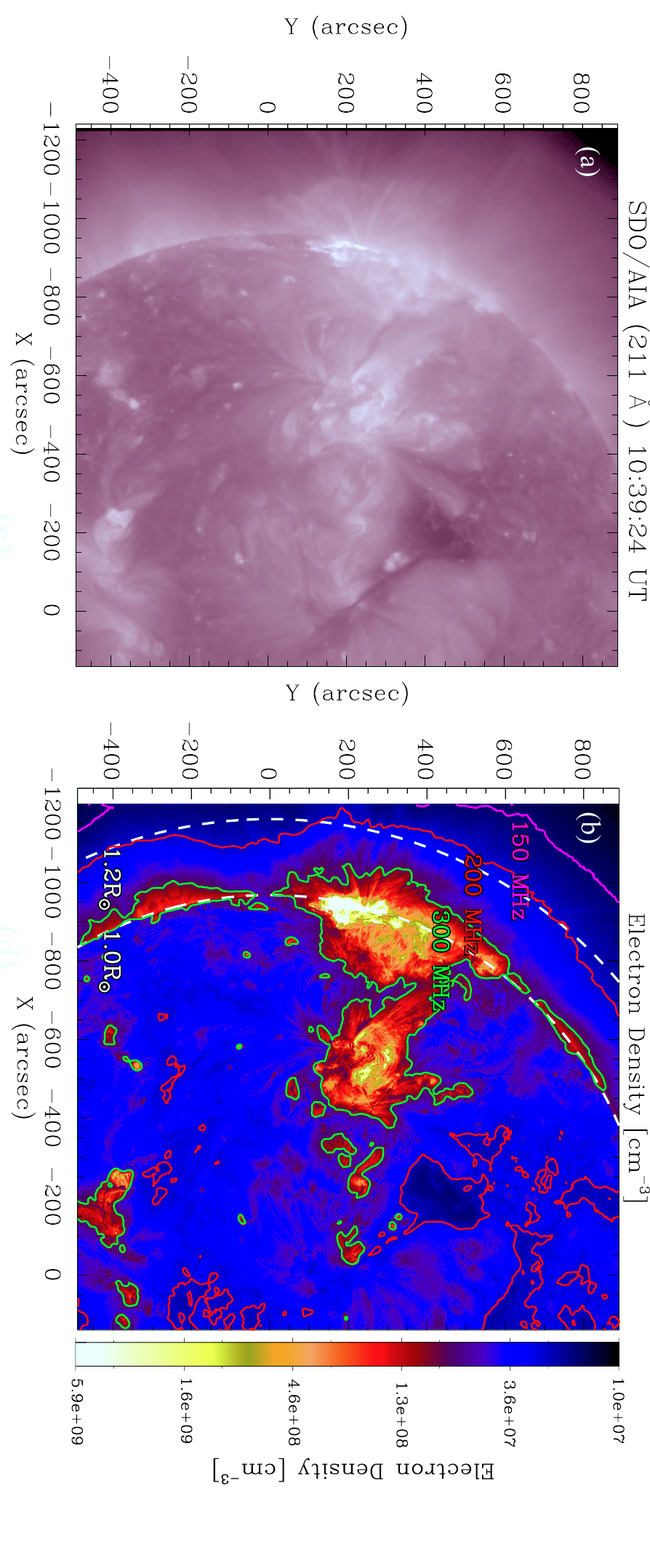}
\caption{{\it SDO}/AIA 211~\AA\ passband intensity at 10:39:24~UT (a) and the corresponding density map (b). The contours show the harmonic plasma emission frequencies related to the measured coronal densities at 150, 200 and 300~MHz. The contours have been smoothed by a boxcar average of 3.6 arcsecs for display purposes, while} dashed lines are at 1.0~$R_{\odot}$ and 1.2~$R_{\odot}$.
\label{Density_map_sdo}
\end{center}
\end{figure*}

Here, a new method to calculate coronal densities, magnetic field strengths, and Alfv\'en speeds in a 2D plane is presented. These 2D maps are then used to calculate the kinematics and the direction of propagation of a Type II radio burst observed on 2011 September 22. The Type II radio burst and coronal mass ejection (CME) are described in Section~\ref{observations}.  The observational method and the models used to produce the 2D density maps are described in Section~\ref{density}, while the Alfv\'en maps are shown in Section~\ref{Alfven}. Results are presented in Section~\ref{results}, and finally, conclusions are discussed in Section~\ref{discussion}.

\section{Observations}
\label{observations}

On 2011 September 22, an X1.4 class flare was observed by \emph{GOES--15}. The flare was identified to have originated in the NOAA active region 11302 and was associated with a CME eruption observed by the Atmospheric Imaging Assembly \citep[AIA;][]{Lemen2012} on board the \emph{Solar Dynamic Observatory} \citep[$SDO$;][]{Pesnell2012} and by the Large Angle and Spectrometric Coronagraph \citep[LASCO;][]{Brueckner1995} on board the \emph{Solar and Heliospheric Observatory} \citep[\emph{SOHO};][]{Domingo1995}. The flare was followed by a Type II radio burst starting  at 10:39:06~UT, which was observed with the e-Callisto spectrometer at the Rosse Solar--Terrestrial Observatory \citep[RSTO;][]{Zucca2012}\footnote{www.rosseobservatory.ie}. 
In Fig.~\ref{RSTO}, the dynamic spectrum from RSTO (20--180~MHz) is shown together with its related \emph{GOES--15} soft X--ray light curve showing the X1.4 flare. The radio burst  shows both fundamental (F) and harmonic (H) emission. The fundamental emission showing a drift rate of $\sim$0.15~MHz~s$^{-1}$ is evident between 35 and 55~MHz, while the harmonic emission lies between 70 and 88~MHz reaching the FM broadcasting radio band. 

The Type II related CME was observed at low altitude with the $SDO$/AIA 211~\AA~(\ion{Fe}{xiv}) bandpass filter. 
Three running difference images are shown in Fig.~\ref{SDO_cme}, which are obtained by subtracting the image recorded one minute prior to the current time (i.e., five 211~\AA~ frames previous). 
The CME leading edge is already outside the $SDO$ field-of-view (FOV) at~10:36:39~UT (Fig.~\ref{SDO_cme}a), while the propagation of the coronal bright front related with the expansion of the CME flanks is visible at 10:38:38~UT (Fig.~\ref{SDO_cme}b) and at 10:40:24~UT (Fig.~\ref{SDO_cme}c).

In Fig.~\ref{Density_map_sdo}a, the intensity from $SDO$/AIA is shown at~10:39:24~UT. The corresponding coronal density map for this field of view (see Section~\ref{sdo_density}), is shown in Fig.~\ref{Density_map_sdo}b. Higher in the corona, electron densities were obtained from $pB$ images from LASCO C2. Only one polarization sequence is taken per day, so 02:57:00~UT was used in this work (see Section~\ref{lasco_density}).

\begin{figure}
\begin{center}
\includegraphics[trim= 5cm 7cm 5cm 7cm,clip=true,width=4.5cm,angle=-90]{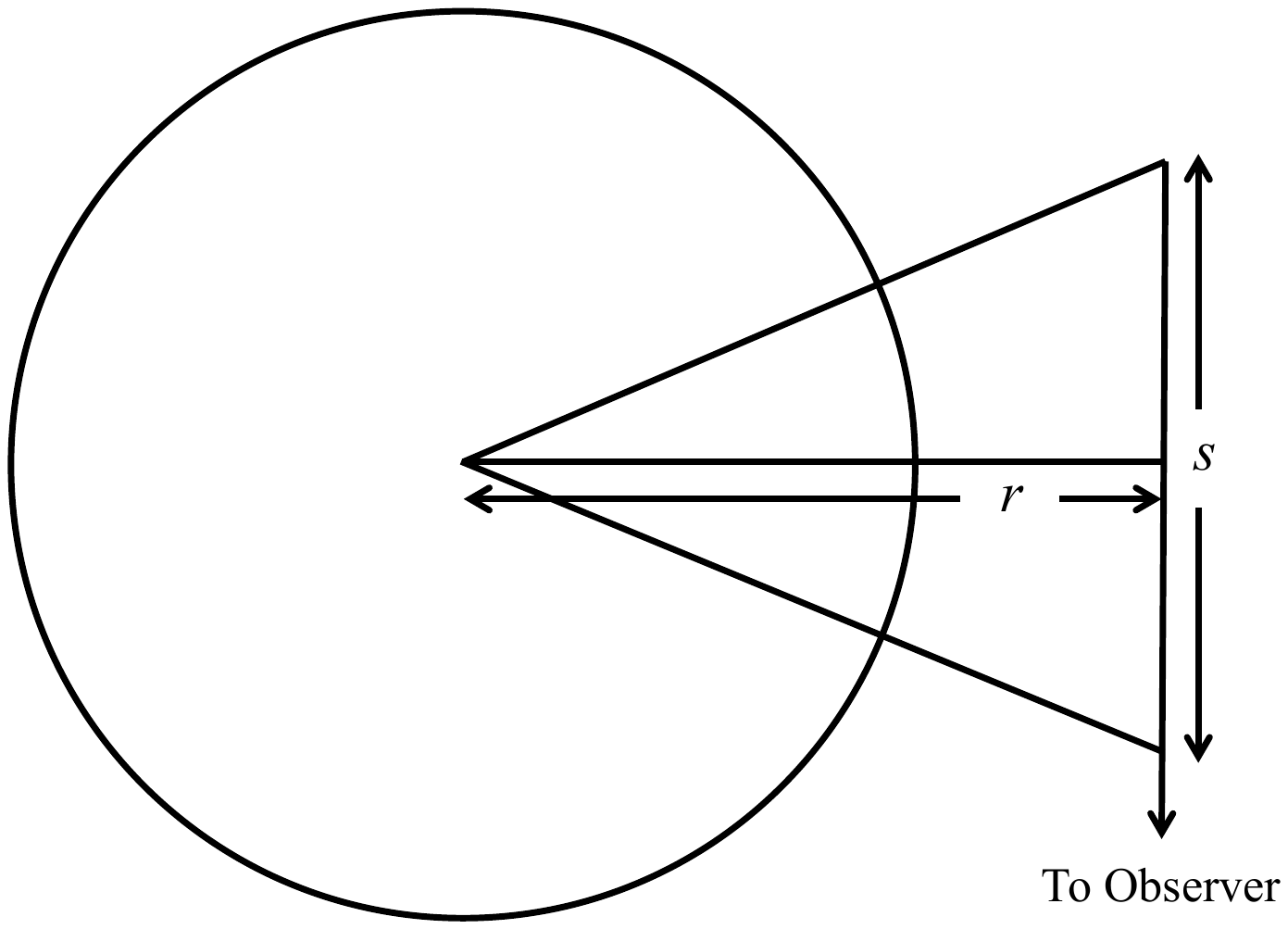}
\caption{The geometry for the emitting solar atmosphere effective LOS path length. The observer measures the contribution of the plasma emitting along the path $s$, which varies with the heliocentric distance $r$.}
\label{LOS_geometry}
\end{center}
\end{figure}
\section{Density maps}
\label{density}
Density was estimated using images from $SDO$/AIA for the height range 1--1.3 $R_{\odot}$ and $SOHO$/LASCO for 2.5--5 $R_{\odot}$. Unfortunately, there are no direct observations available at 1.3--2.5~$R_{\odot}$. As a result, a data-constrained density model was used to interpolate between the $SDO$/AIA and $SOHO$/LASCO data coverage.
In the following subsections the calculation of this map is described for the $SDO$/AIA density, the $SOHO$/LASCO density, and finally, the data-constrained analytic model.

\subsection{$SDO$/AIA densities ($<1.3~R_{\odot}$)}
\label{sdo_density}
Electron densities were calculated from emission measure maps derived using the $SDO$/AIA's six coronal filters and the method of \citet{Aschwanden2011}. The method starts by reconstructing the differential emission measure $dEM/dT$ (DEM) using the intensity of the six $SDO$/AIA filters for each pixel. The DEM is a measure of the amount of plasma along the line-of-sight (LOS) that contributes to the emitted radiation in the temperature range $T$ to $T+dT$ \citep{Craig1976}. 
Once the column $EM$ was obtained by integrating the DEM over the temperature range $dT$, the plasma electron density can be calculated by estimating an effective path length of the emitting plasma along the LOS. The 2D $EM(r,\phi)$ map, which is a function of heliocentric distance $r$ and latitude $\phi$, can then be written as,
\begin{align}
\label{Emission_measure}
EM(r,\phi) & = \int{\left ( \frac{dEM(r,\phi,T)}{dT} \right )dT}, \nonumber \\ 
& = \int <N_\mathrm{e}^{2}(r,\phi)> ds \mathrm{~~~~~~[cm^{-5}]}.
\end{align}
Knowing the effective LOS path length, $s$, the density of the emitting plasma can be obtained from the $EM$,
\begin{equation}
N_\mathrm{e}(r,\phi)=\sqrt{\frac{EM(r,\phi)}{s(r)}} \mathrm{~~~~~~[cm^{-3}]}.
\end{equation}
The effective LOS path length was calculated using a geometrical method used widely in stellar atmospheres \citep{Menzel1936} with a schematic of the problem shown in Fig.~\ref{LOS_geometry}.   The length of $s$ changes at different heliocentric distances $r$, which contributes to the intensity of the emitting plasma measured by the observer. 
This gives the effective LOS path length using an asymptotic series expansion in the form,
\begin{equation}
s \sim (H \pi r)^{1/2},
\end{equation}
where $H$ is the scale height. 
Using a typical coronal temperature of 2~MK, the scale height $H$ measures $\sim9\times10^9$ cm and $s$ $\sim4\times10^{10}$ cm. The value of $s$ does not significantly change in the 1--1.3 $R_{\odot}$ range. 
The density map obtained with $SDO$/AIA centered on NOAA 11302, and the related harmonic plasma frequency contours at 150, 200, and 300~MHz are shown in Fig.~\ref{Density_map_sdo}b.

\begin{figure}
\begin{center}
\includegraphics[trim=2cm 2cm 2cm 3.5cm,clip=true,width=7cm,angle=90]{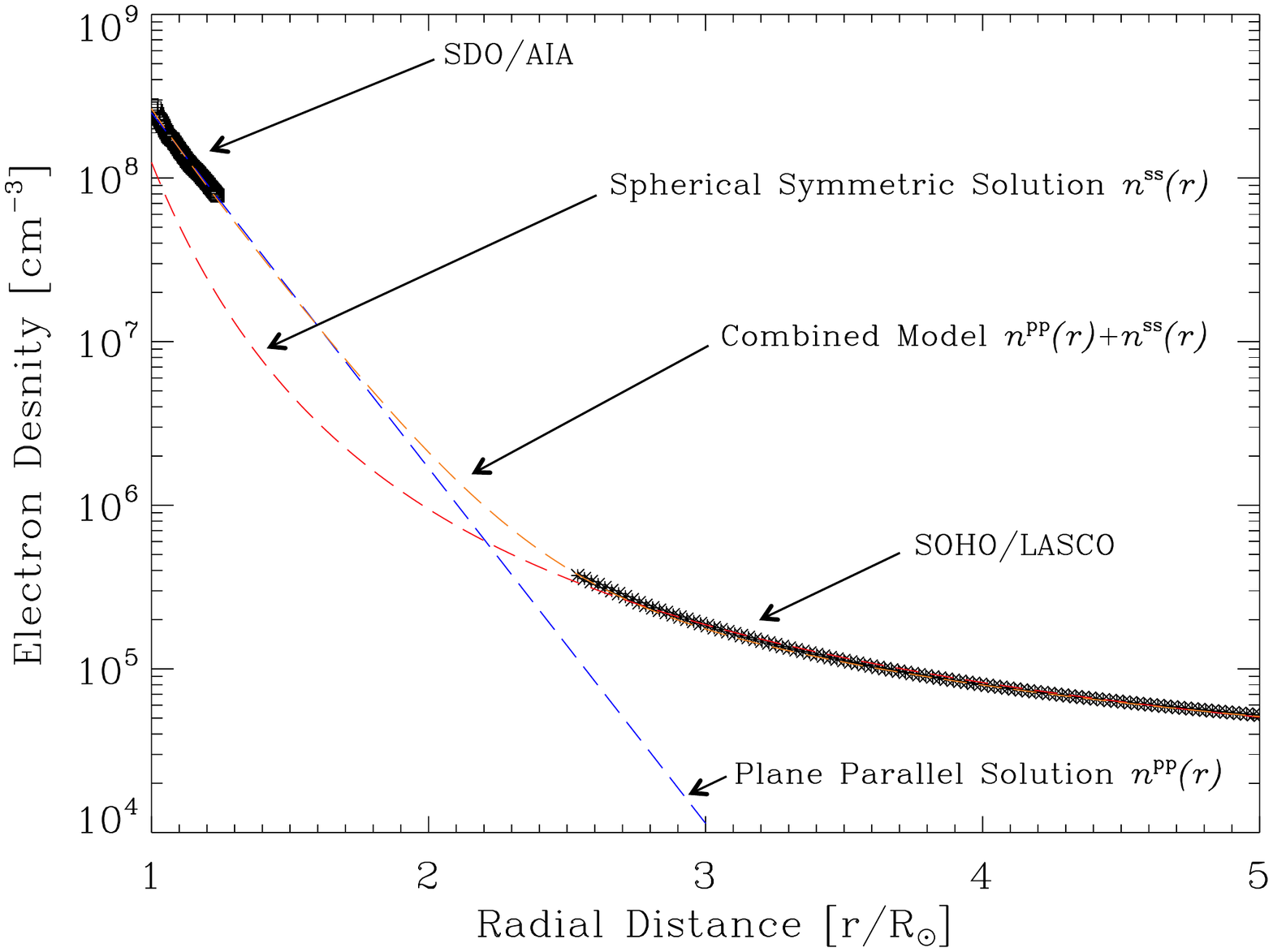}
\caption{A radial profile of the coronal density model constrained by densities from $SDO$/AIA and $SOHO$/LASCO. In blue, the plane-parallel solution reproduces the active region, (Eq.~\ref{planeparallel_hydrostatic}), and in red, the spherical-symmetric solution reproduces the quiet Sun (Eq.~\ref{spherical_hydrostatic}). The combined model in orange (Eq.~\ref{planeparallel_spherical_hydrostatic}) is used to replace the missing data in the range 1.3-2.5~$R_{\odot}$.}
\label{Density_models_fit}
\end{center}
\end{figure}

\subsection{$SOHO$/LASCO densities ($>2.5~R_{\odot}$)}
\label{lasco_density}
For the density calculation using $SOHO$/LASCO, we use $pB$ images. The K-coronal brightness results from Thomson scattering of photospheric light by coronal electrons \citep{Billings1966}. The intensity of scattered light and its polarization depends on the number of scattering electrons and a number of geometric factors, which was first outlined by \citet{Minnaert1930}. A method for estimating the electron density using these geometric factors and polarized brightness observations was first employed by  \citet{Hulst1947}, which remains the standard procedure today. 
The F corona (arising from interplanetary dust scattering) must be eliminated from the data. In the case of $pB$ observations at small elongations \citep[$\leq$5~$R_{\odot}$;][]{Mann.I1992}, the F corona can be assumed unpolarized and thus does not contribute to the $pB$ signal; hence, we restrict our analysis to $\leq5~R_{\odot}$. For full details on the calculation see \citet{Hayes2001}.



\begin{figure}
\begin{center}
\includegraphics[trim=3cm 4cm 4cm 5cm,clip=true,width=7cm,angle=90]{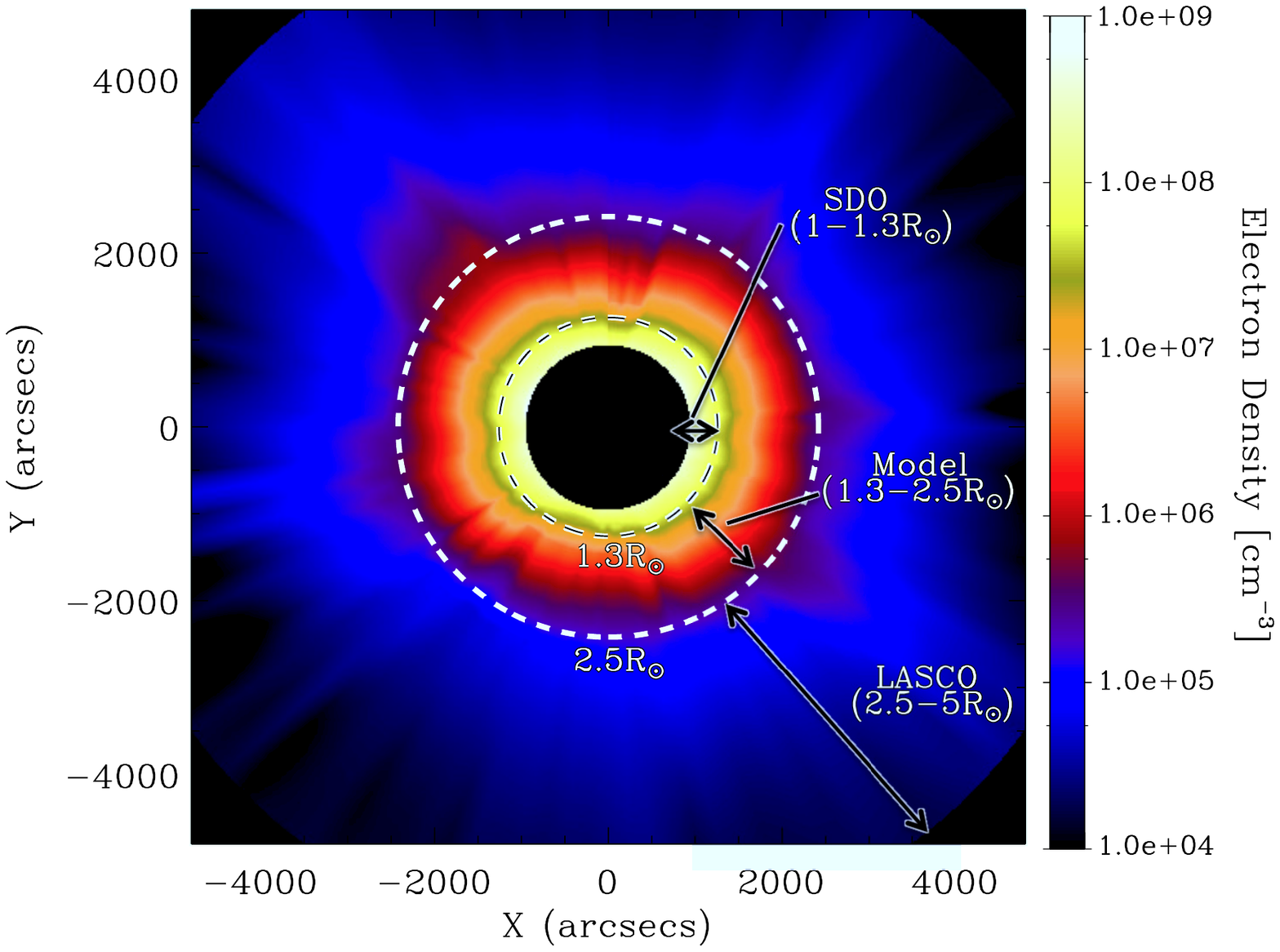}
\caption{The 2D electron density map on 2011 September 22 obtained from $SDO$/AIA (1--1.3~$R_{\odot}$) and $SOHO$/LASCO (2.5--5~$R_{\odot}$) with an interpolated electron density from the density model given in Eq.~\ref{planeparallel_spherical_hydrostatic} (1.3--2.5~$R_{\odot}$).}
\label{2D_density}
\end{center}
\end{figure}

\subsection{Data-constrained density model ($1.3-2.5~R_{\odot}$)}
\label{models}
For the height range 1.3--2.5~$R_{\odot}$, a combined plane-parallel and spherically-symmetric model, which is described below, was employed. Assuming a spherically-symmetric corona in hydrostatic equilibrium, the gradient of the pressure, $P$, is balanced by gravity,

\begin{equation}
\frac{dP(r)}{dr}=-\rho(r) g(r),
\end{equation}
where $\rho(r)$ is the coronal plasma mass density and $g( r ) = GM_{\odot} / r^2$. By integrating, using the ideal gas law ($P = 2nkT$), the spherically--symmetric (SS) density, $n^{\mathrm{ss}}(r)$,  in an hydrostatic stratified corona is derived as,
\begin{equation}
\label{spherical_hydrostatic}
n^{\mathrm{ss}}(r)=n^{\mathrm{ss}}_0\exp\left[-\frac{1}{H} \left(r-R_{\odot}\right) \right],
\end{equation}
where $n^{\mathrm{ss}}_0$ is the SS electron density at the solar surface (i.e., $r=1~R_{\odot}$), $H(r)=kTr^2/\mu m_\mathrm{p}GM_{\odot} $ is the scale height, $k$ is the Boltzmann constant, $T$ is the plasma temperature, $\mu$ is the mean molecular weight, $m_\mathrm{p}$ is the proton mass, $G$ is the gravitational constant and $M_{\odot}$ is the solar mass. This expression holds well for quiet Sun conditions at large distances from the Sun (i.e., $r$\textgreater$2~R_{\odot}$). 

Low in the solar atmosphere, where $r$\textless\textless$2~R_{\odot}$, this reduces to the plane--parallel (PP) solution,
\begin{equation}
\label{planeparallel_hydrostatic}
n^{\mathrm{pp}}(r)=n^{\mathrm{pp}}_0\exp\left[-\frac{1}{H_0}(r-R_{\odot}) \right],
\end{equation}
where $n^{\mathrm{pp}}_{0}$ is the PP electron density, $H_{0} = k T /  \mu m_{p} g_{\sun}$ is the scale height, and $g_{\sun}$ is the acceleration due to gravity, which are all defined at the solar surface (i.e., $r = 1 R_{\sun}$).The PP solution is a good approximation of the electron density distribution in the low corona and in active regions \citep[see,][]{Aschwanden2001}.

\begin{figure}
\begin{center}
\includegraphics[trim=0cm 0cm 1cm 1.5cm,clip=true,width=6cm,angle=90]{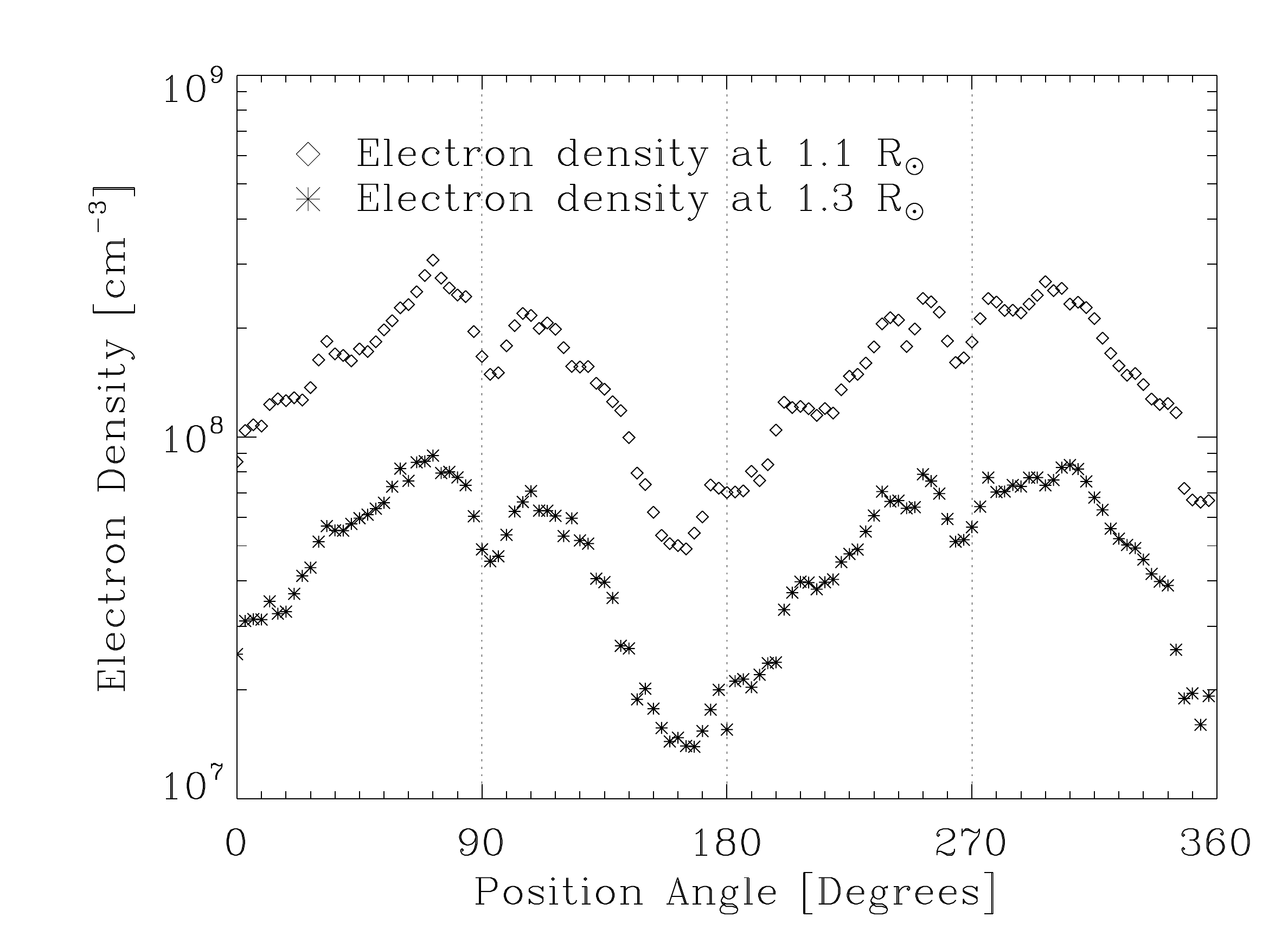}
\caption{Electron density at different position angles (PA) for heights of 1.1~$R_{\odot}$ and 1.3~$R_{\odot}$. Coronal holes measure a density of $\sim$$3\times10^7$~cm$^{-3}$ at 1.1 $R_{\odot}$ and $\sim$$1\times10^7$~cm$^{-3}$ at 1.3 $R_{\odot}$, while the active regions reach a density of $\sim$$2\times10^8$~cm$^{-3}$ at 1.1 $R_{\odot}$ and $\sim$$9\times10^7$~cm$^{-3}$ at 1.3 $R_{\odot}$.}
\label{polar_density}
\end{center}
\end{figure}

In this work, we simultaneously model the outer corona using the SS model ($r$\textgreater$2~R_{\odot}$) and the possible presence of a PP active region at $r$\textless$2~R_{\odot}$ using,
\begin{equation}
\label{planeparallel_spherical_hydrostatic}
n(r)=n^{\mathrm{pp}}(r)+n^{\mathrm{ss}}(r).
\end{equation}
This was then used to interpolate the observational data between 1.3~$R_{\odot}$ and 2.5~$R_{\odot}$.
Figure~\ref{Density_models_fit} presents an example electron density radial profile with data from $SDO$/AIA and $SOHO$/LASCO. The PP solution (Eq.~\ref{planeparallel_hydrostatic}) is displayed in blue. It reproduces the density enhancement of an active region well,  but it decays too fast to reproduce the quiet Sun (QS) density at large coronal heights. The SS solution (Eq.~\ref{spherical_hydrostatic}) is displayed in red and it reproduces the quiet Sun coronal density. The combined PP and SS model (Eq.~\ref{planeparallel_spherical_hydrostatic}) is displayed in orange. The combined model fits the observational data well and is employed in producing the density map for the height range 1.3--2.5 $R_{\odot}$, which is shown in Fig.~\ref{2D_density}. The map shows the difference in electron density between polar and equatorial regions and the presence of coronal streamers and active regions are clear. Electron density at different position angles for the heights of 1.1 and 1.3 $R_{\odot}$ are shown in Fig.~\ref{polar_density}. Coronal holes (CH) have a density of $\sim3\times10^7$ electrons cm$^{-3}$ at 1.1 $R_{\odot}$ at the position angle of $\sim$170\degree\ and $\sim1\times10^7$ cm$^{-3}$ at 1.3 $R_{\odot}$. Meanwhile, active regions (AR) reach a density of $\sim2\times10^8$ cm$^{-3}$ at 1.1 $R_{\odot}$ at a position angle of $\sim$ 70\degree\ and $\sim9\times10^7$ cm$^{-3}$ at 1.3 $R_{\odot}$. These density maps are in good agreement with those derived using EUV line ratios by \citet{Gallagher1999}.

\section{Alfv\'en speed maps}
\label{Alfven}
The Alfv\'en speed, which is the speed at which information travels in a magnetized plasma, was obtained by combining measurements of electron density (described in Section~\ref{density}) and magnetic field strength. Specifically, the 2D Alfv\'en speed map was calculated using a 2D magnetic field plane obtained from 3D magnetic extrapolations, which are described in the following section. 
\subsection{Magnetic field}

The coronal magnetic field strength was obtained from the potential-field source-surface (PFSS) extrapolation, which is based on \citet{Schatten1969} and the software of  \citet{Schrijver2003}\footnote{http://www.lmsal.com/$\sim$derosa/pfsspack/}. The PFSS package combines measured LOS photospheric magnetograms with an evolving surface flux transport model. This provides full solar surface coverage by evolving magnetic flux that rotates over the Earth-viewed western limb to cover the far side  of the Sun. This spherical surface is then used as the lower boundary condition for the PFSS extrapolation, which  provides a vector magnetic field solution for a 3D grid of polar coordinates ($r$ - radial distance; $\theta$ - longitude; $\phi$ - latitude). The vector field is described by components in the direction of each of the polar coordinates (i.e., $B_{r}$, $B_{\theta}$, $B_{\phi}$), such that the total magnetic field, $B$, is given by  
\begin{figure}
\begin{center}
\includegraphics[trim=7cm 6cm 4.2cm 4.2cm,clip=true,width=11cm,angle=0]{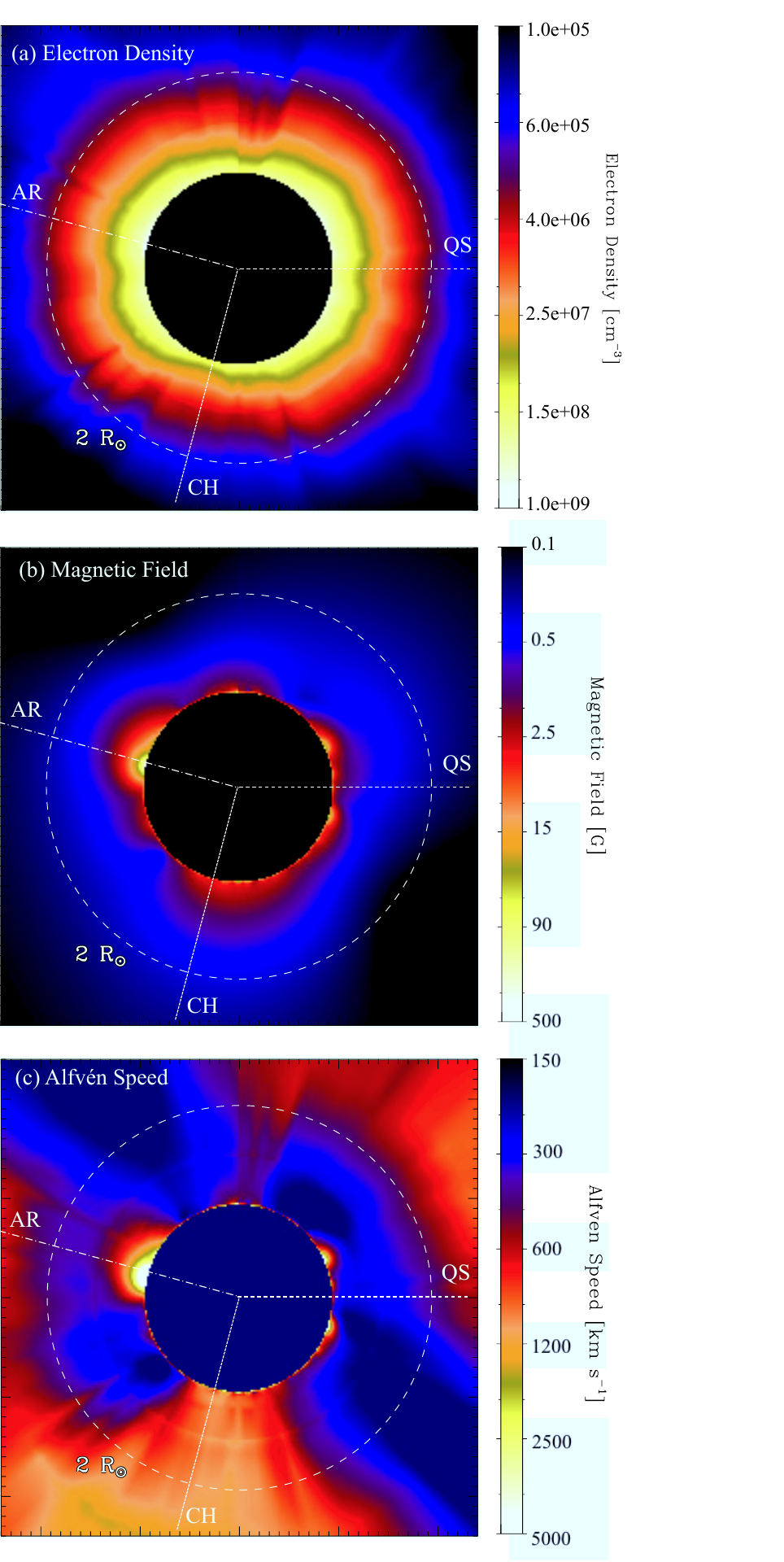}
\caption{(a) 2D electron density map from 2011 September 22. (b) The magnetic field strength obtained with PFSS extrapolation, and (c) the Alfv\'en speed map obtained from the electron density and magnetic field strength values (Eq.~\ref{alfven_equation}). The active region (AR), quiet Sun (QS), and coronal hole (CH) radial profiles used in Fig.~\ref{density_3profiles} are also shown.}
\label{2D_density_bfield_alfven}
\end{center}
\end{figure}
\begin{figure}
\begin{center}
\includegraphics[trim=1.5cm 3cm 1cm 2.5cm,clip=true,width=9cm,angle=0]{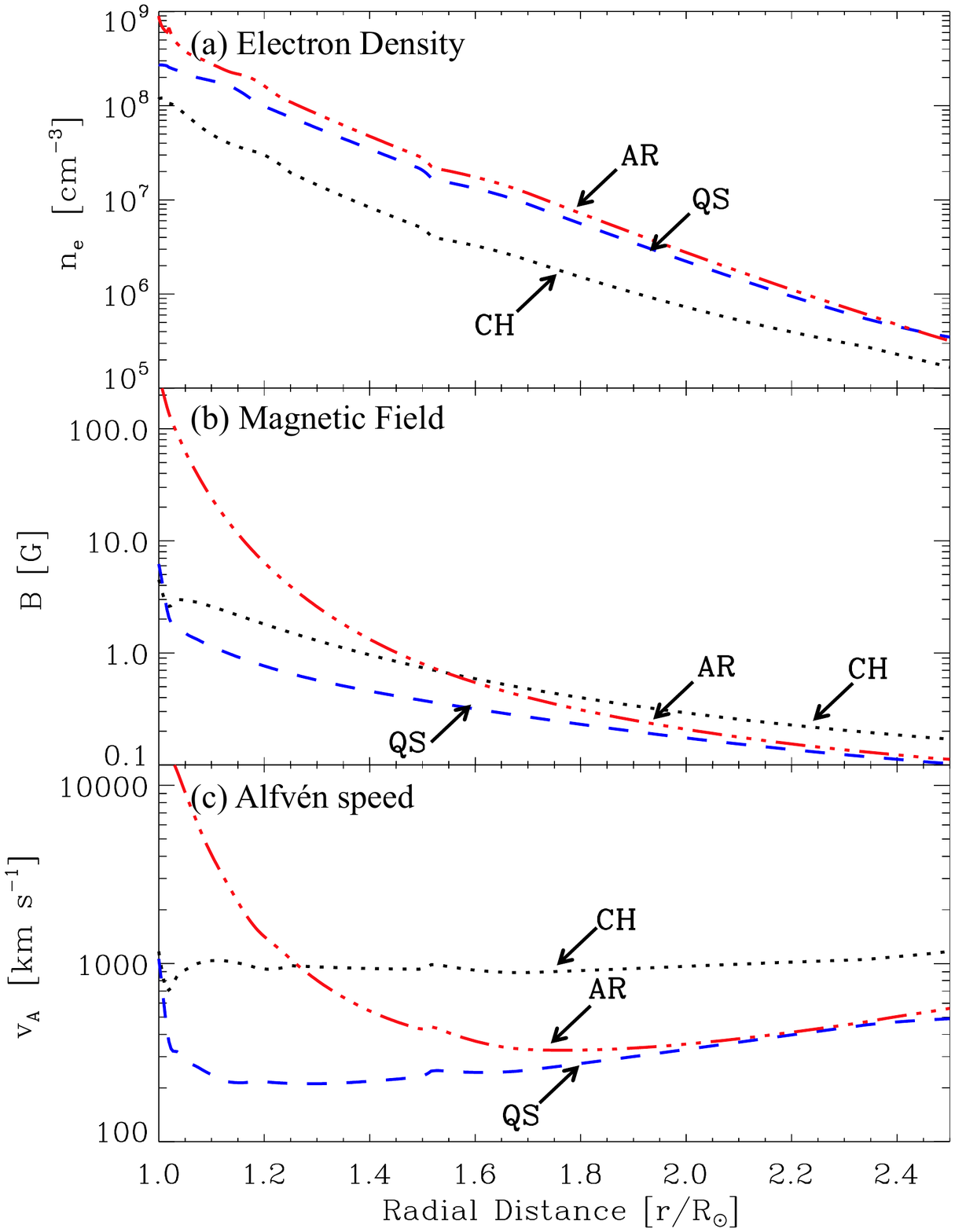}
\caption{(a) The electron density at a function of height for an active region (PA=75\degree) the quiet Sun (PA=270\degree) and coronal hole(PA=165\degree). (b) The magnetic field strength is shown for the same position angles.(c) The calculated Alfv\'en speed for the three profiles. Active region (AR), coronal hole (CH), and quiet Sun (QS)} profiles are displayed in Fig.~\ref{2D_density_bfield_alfven}.
\label{density_3profiles}
\end{center}
\end{figure}

\begin{equation}
B(r,\theta,\phi) = \sqrt{ {B_{r}}^{2} + {B_{\theta}}^{2} + {B_{\phi}}^{2}}.
\end{equation}

One limitation to this model is the handling of magnetic field that emerges on the far side of the Sun. Such fields may not be fully present in the PFSS model for several days after their location rotates over the Earth-viewed eastern limb. This is due to the gradual transition from only flux-transported fields at and behind the eastern limb to only measured fields some distance onto the visible disk.

At the time of the event studied here (2011 September 22 10:39 UT), the photospheric signature of NOAA 11302 was not present in the PFSS model due to its proximity to the east limb (heliographic coordinates N11E81). This region first appears in the PFSS lower boundary on 2011 September 24 but shows significant flux imbalance until 2011 September 26. The PFSS model used here was taken from 2011 September 26 at 12:04 UT, and the effective Earth-viewed plane--of--sky (POS) from 2011 September 22 was extracted. This was achieved by averaging $B(r,\theta,\phi)$ over a \textpm 10\degree\ range of longitudes, $\theta$, centered on both the AR Carrington longitude and its 180\degree-separated location; the result is shown in Fig.~\ref{2D_density_bfield_alfven}b.

\begin{figure}
\centering
\includegraphics[trim=9cm 2.8cm 6.3cm 2.2cm,clip=true, width=8.7cm, angle=0]{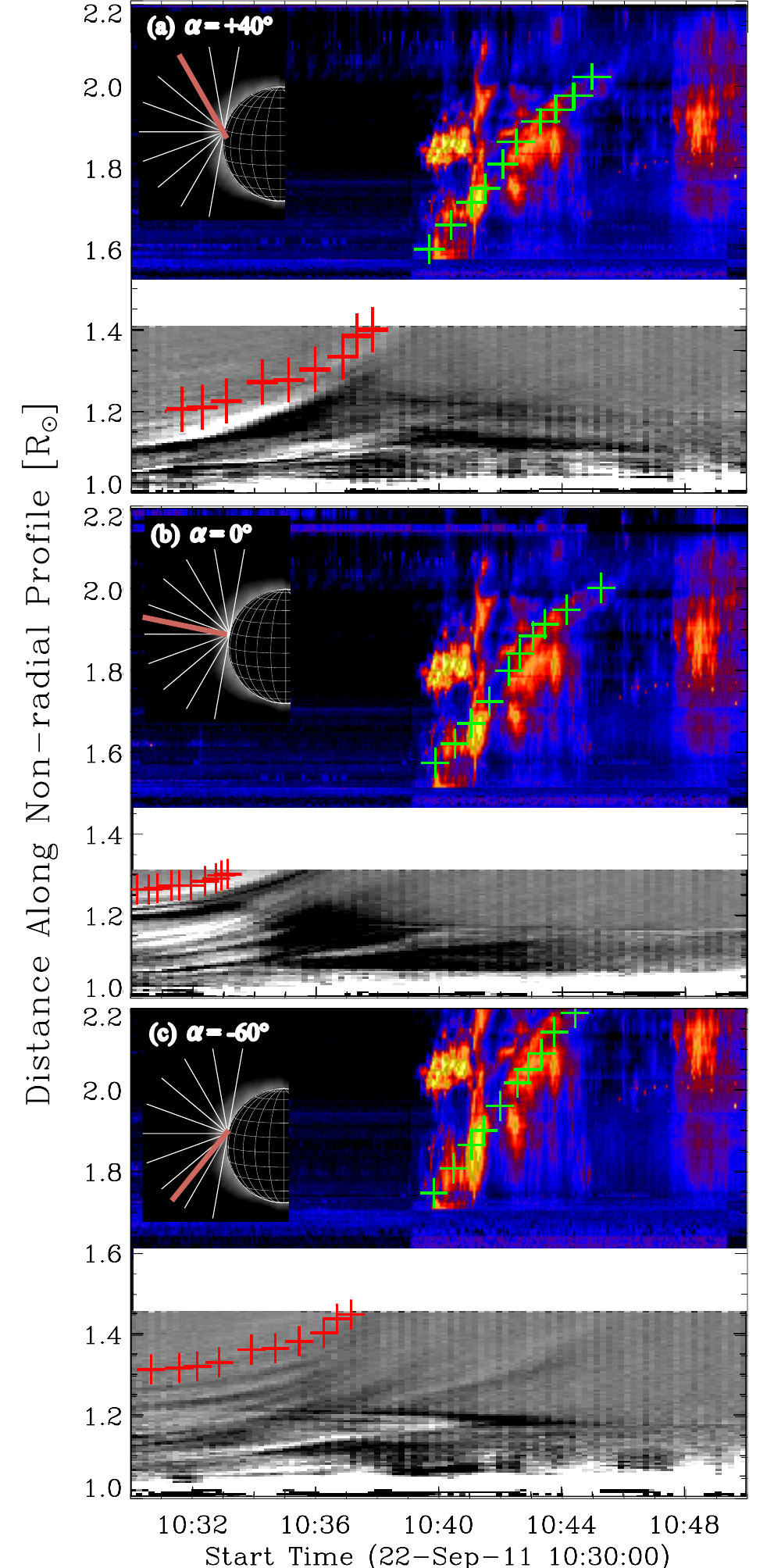}
\caption{Distance-time plot at different propagation angles of the CME observed with $SDO$/AIA running difference images (red data points) and of the Type II radio burst using different non-radial density profiles (green data points). The SDO distance-time plot was obtained by plotting the intensity of the running difference image (grey scale) for the specific trace over time. The Type II distance-time was obtained by plotting the dynamic spectra with the frequency axis converted in height using the density of the correspondent non-radial trace for the harmonic emission. The considered profiles on top of the active region is shown on the inset of each plot. The profiles are marked with white traces separated by 10\degree; the profile used is marked with a red trace. (a) The distance-time plot for the northern flank region at +40\degree\ from the radial trace, and (b) distance-time for the Radial profile (0\degree),  while the distance-time for the southern flank region at -60\degree\ from the radial trace is in panel (c).}
\label{Height_time_euv_typeII}
\end{figure}

\subsection{Alfv\'en speed}

The 2D map of Alfv\'en speed, $v_\mathrm{A}(r,\phi)$, is obtained using the electron density and the magnetic field maps by
\begin{equation}
\label{alfven_equation}
v_\mathrm{A}(r,\phi)=\frac{B(r,\phi)}{\sqrt{4\pi n_\mathrm{e}(r,\phi)\mu m_\mathrm{p}}}.
\end{equation}
Figures.~\ref{2D_density_bfield_alfven}a and \ref{2D_density_bfield_alfven}b present the 2D maps of density and magnetic field used to calculate the Alfv\'en speed. At the base of the corona, the extrapolated magnetic field reaches $\sim$500\,G in active regions, while it reaches $\sim$3-10\,G outside of active regions (i.e., across both quiet Sun and coronal hole regions). The resulting 2D Alfv\'en speed map is shown in Fig.~\ref{2D_density_bfield_alfven}c. The Alfv\'en speed reaches $\sim$$10^4$ km~s$^{-1}$ in active regions and decreases to $\sim$200 km s$^{-1}$ in local minima in neighboring QS regions. Also notable are the relatively high Alfv\'en speeds ($\sim$1000 km s$^{-1}$) in coronal holes because the magnetic field in CHs decreases more slowly than that in ARs and the electron density is significantly lower in CHs than in ARs.  Radial profiles of density, magnetic field, and Alfv\'en speed are shown in Figs.~\ref{density_3profiles}a, \ref{density_3profiles}b, and \ref{density_3profiles}c, respectively, for an active region (PA=75\degree), quiet Sun (PA=270\degree) and coronal hole regions (PA=165\degree). 
\section{Results and discussion}
\label{results}

A CME was observed low in the corona with $SDO$/AIA (Fig.~\ref{SDO_cme}), which is the candidate for triggering the Type II radio burst observed with RSTO (Fig.~\ref{RSTO}). To verify this, the kinematics of the Type II burst and the CME were calculated. The kinematics of the Type II radio burst were calculated relating the plasma frequency measured in the dynamic spectrum to its electron density ($f_p\propto\sqrt{n_e} $). The position of the shock is then obtained by relating the emitting electron plasma density to its position on the calculated density map (Fig.~\ref{2D_density}). To date, this has been done by employing radial models of the coronal density even if Type II radio bursts propagate often non-radially.    

Using the calculated 2D density map, non-radial density traces from NOAA 11302 were extracted. The traces, which start at the solar region from the solar limb, have 10 degree separation from the radial trace $(\alpha=0\degree)$ with positive values toward solar north and negative values toward solar south. For each of the non-radial profiles a distance--time plot for the CME and the Type II radio burst was obtained.
\begin{figure}
\centering
\includegraphics[trim=9cm 5.7cm 9cm 1.3cm,clip=true, width=9cm, angle=0]{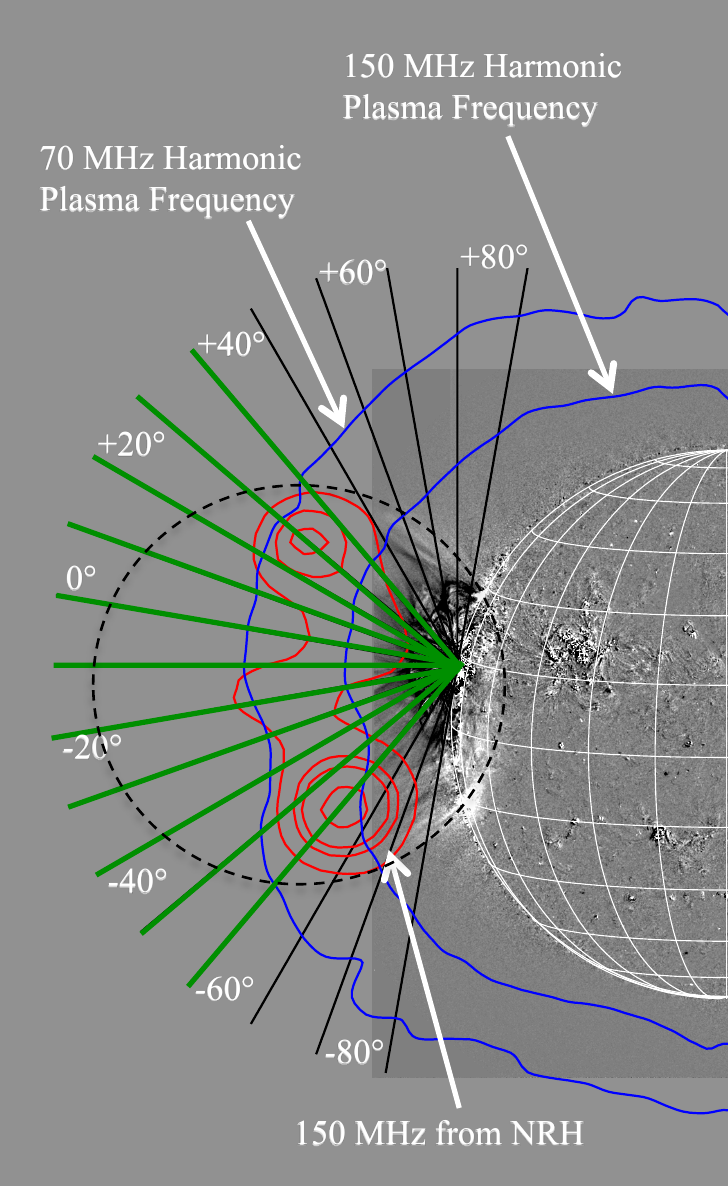}
\caption{Running difference image of the CME observed with $SDO$/AIA 211 \AA\ passband at 10:40:24 UT with superimposed, non--radial profiles used to calculate the Type II radio burst and CME kinematics (black traces). Traces showing matching kinematics between the CME and the Type II burst are marked in green. The red contours are the Nan\c cay NRH emission at 150MHz at 10:40:26 UT. The blue contours are the harmonic plasma emission at 70 and 150 MHz estimated from the density map. The suggested locations of the Type II burst at the CME flanks are positioned at the intersection between the CME front (indicated with the dashed black curve) and the 70 MHz harmonic plasma frequency contour (in blue).}
\label{profiles_nancay}
\end{figure}
The CME kinematics were obtained using $SDO$/AIA 211~\AA~ running difference images. Three example frames of running difference images are shown in Fig.~\ref{SDO_cme}. For each of the non--radial traces, a distance--time plot was obtained. In Fig.~\ref{Height_time_euv_typeII}, three example distance--time plots are shown for $\alpha=+40\degree$,$0\degree$,$-60\degree$. The CME distance-time evolution obtained from the running difference intensity along the non-radial trace is displayed in gray scale with explicit data points in red. The Type II radio burst dynamic spectra are displayed in red-blue color scale, where frequency has been converted into distance along the non-radial trace. This was achieved using the density obtained along that trace and by considering a harmonic plasma emission. The data points related with the position of the harmonic emission are displayed in green. The panel at the left of each plot indicates the considered non-radial trace.
Comparable kinematics were found between the $SDO$/AIA front and the Type II radio burst position for $\alpha$ between $-60\degree$ and $+40\degree$ (green traces in Fig.~\ref{profiles_nancay}). We note that the SDO FOV limits the maximum observable height to $\sim$1.3~$R_{\odot}$ for near-radial traces (e.g., $\alpha$ between $-20\degree$ and $+20\degree$), making the kinematic comparison with the Type II radio burst complicated. In addition, the SDO running difference front for these traces (see Fig.~\ref{Height_time_euv_typeII}a) is most likely related with the loop structure emerging below the CME leading edge, which is already outside the FOV of SDO (see Fig.~\ref{SDO_cme}).

In Fig.~\ref{profiles_nancay}, the $SDO$/AIA 211~\AA\ running difference image shows the lower portion of the CME at 10:40:24 UT with the Nan\c cay radio heliograph \citep[NRH;][]{Kerdraon1997} emission at 150~MHz, which is superimposed as red contours (40\%, 60\%, 70\%, and 90\% of the peak flux). The NRH emission at 150~MHz shows a brightness temperature of $log_{10}T=7.1$, while the dynamic spectrum at 150~MHz shows a faint continuum (Fig.~\ref{RSTO}). This emission is most likely originated from the CME plasma \citep[see,][]{Ramesh2005}. The relation of the 150~MHz emission and the CME is also evident from the spatial correspondence of the NRH contours and the CME flanks.
The harmonic plasma emission contours at 70 and 150~MHz were calculated from the density map are displayed in blue. The 70 MHz contour corresponds to the frequency of the Type II shock upstream region (harmonic emission in Fig.~\ref{RSTO}). The 150 MHz contour is shown for height comparison with the NRH emission. We suggest two locations where the Type II radio burst could be originated; these positions at the CME flanks are located where the 70~MHz contour (Type II upstream region) intersects the CME front edge (indicated with a black dashed line) at a height of $\sim$1.6~$R_{\odot}$. 
For the same CME eruption, \citet{Carley2013} found that a series of herringbone radio bursts have originated along the southern flank as it expands through the corona. This agrees with our finding of comparable kinematics between the Type II shock and the CME expansion in the southern flank region (i.e., $\alpha=-60\degree$). 

\begin{figure*}
\begin{center}
\includegraphics[trim=4.5cm 2.6cm 7.5cm 3cm,clip=true,width=7cm,angle=-90]{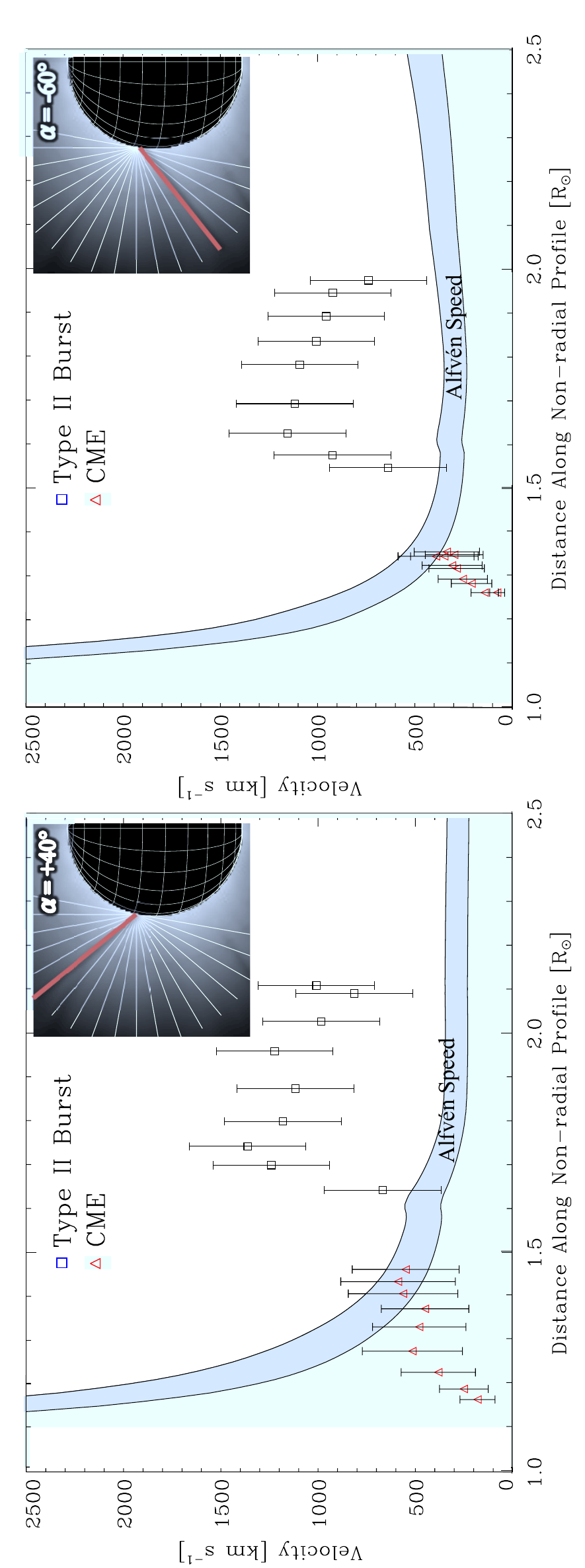}
\caption{Propagation speed calculated for the CME and the Type II radio burst for two different non--radial profiles in the \textquotedblleft flank\textquotedblright~region at $\alpha=+40\degree$ and $\alpha=-60\degree$ (profiles are displayed on the insets). The calculated Alfv\'en speed is displayed in the shaded region.}
\label{Alfven_speed}
\end{center}
\end{figure*}

The speeds of the CME and of the Type II radio burst for the flank regions ($\alpha=+40\degree$, $-60\degree$) are compared with the Alfv\'en speed (shaded area) in Fig.~\ref{Alfven_speed}. The Alfv\'en speed was extracted along the non-radial traces ($\alpha=+40\degree$, $-60\degree$) from the 2D map shown in Fig.~\ref{2D_density_bfield_alfven}c. The CME flank speed remains sub-Alfv\'enic, while the Type II burst shows a super-Alfv\'enic speed. The Type II burst in the flank regions reaches a max speed of $\sim$$1400$ km~s$^{-1}$ and then decelerates to $\sim700$ km~s$^{-1}$. This deceleration may result in the fading of the Type II burst in the dynamic spectra as the burst speed approaches the local Alfv\'en speed.

\section{Conclusion}
\label{discussion}

A new method to obtain semi--empirical 2D maps of coronal density and Alfv\'en speed has been presented. Using these calculated maps, a decametric Type II burst and a CME that occurs on 2011 September 22, were analyzed to understand their relationship.

The analysis of the Type II kinematics, using the 2D density maps, have been performed for non--radial profiles in the POS starting from the active region. Previously, analytical radial density models have been employed to calculate the radio burst kinematics \citep[e.g., ][]{Robinson1985,Vrsnak2002}. Using a 2D density map at the specific time of the radio burst  and performing a non--radial analysis, we were able to relate the Type II burst with the CME and discriminate which portion of the CME front was responsible for triggering the Type II emission. Comparing the CME kinematics calculated for different non-radial traces to the Type II burst we find evidence for the shock to be formed in the flank region of the CME. This agrees with previous interpretations \citep[e.g.,][]{Classen2002,Mancuso2004}. By calculating the height of the upstream shock front (70~MHz) using the 2D density map and relating it to the position of the CME, we were able to locate the position of the shock in the CME flanks at a height of $\sim$1.6~$R_{\odot}$.
 
In addition, we were able to compare the Type II burst shock speed with the local Alfv\'en speed using the 2D Alfv\'en speed maps . The Type II burst speed was found to be super-Alfv\'enic \citep[e.g.,][]{Vrsnak2008,Cho2005,Klein1999}). Furthermore, the decelerating phase of the Type II shock can be related with the fading of the Type II burst in the dynamic spectrum as the shock approaches the local Alfv\'en speed.

The method used to construct 2D density and Alfv\'en speed maps presented in this work can be used for the study of all Type II radio bursts. These time specific semi-empirical maps improve the calculation of shock kinematics and represent a step forward from the standardized radial density profiles currently employed. Further investigation relating the kinematics of the limb event of Type IIs and CMEs using the 2D maps and radio imaging is necessary. This can be done with Type II radio bursts in the NRH frequency range (i.e., 150--445~MHz) where direct radio imaging can be used to locate the position of the burst, or with lower frequency imaging telescopes, such as the LOw Frequency ARray \citep[LOFAR;][]{VanHaarlem2013}.

\begin{acknowledgements}

PZ is supported by a TCD Innovation Bursary. EC is a Government of Ireland Scholar supported by the Irish Research Council. DSB is supported by the European Space Agency Prodex programme. We are grateful to the SDO, LASCO and Nan\c cay NRH team for the open access data. We are grateful to David Long for the beneficial discussions and help.

\end{acknowledgements}

\email{zuccap@tcd.ie}.

\clearpage


\begin{thebibliography}{}

\bibitem[Allen(1947)]{Allen1947} Allen, C.~W.\ 1947, \mnras, 107, 426 

\bibitem[Aschwanden et al.(2001)]{Aschwanden2001} Aschwanden, M.~J.,Schrijver, C.~J., \& Alexander, D.\ 2001, \apj, 550, 1036 

\bibitem[Aschwanden et al.(2011)]{Aschwanden2011} Aschwanden, M.~J., Boerner, P., Schrijver, C.~J., \& Malanushenko, A.\ 2011, \solphys, 384 

\bibitem[Bemporad et al.(2003)]{Bemporad2003} Bemporad, A., Poletto, G., \& Romoli, M.\ 2003, \memsai, 74, 721 

\bibitem[Bougeret et al.(2008)]{Bougeret2008} Bougeret, J.~L., Goetz, K., Kaiser, M.~L., et al.\ 2008, \ssr, 136, 487 

\bibitem[Billings(1966)]{Billings1966} Billings, D.~E.\ 1966, New York: Academic Press, |c1966,  

\bibitem[Brueckner et al.(1995)]{Brueckner1995} Brueckner, G.~E., Howard, R.~A., Koomen, M.~J., et al.\ 1995, \solphys, 162, 357 

\bibitem[Cane et al.(1981)]{Cane1981} Cane, H.~V., Stone, R.~G., Fainberg, J., et al.\ 1981, \grl, 8, 1285 

\bibitem[Carley et al. (2013)]{Carley2013} Carley, E. P., Long, D. M., Byrne, J. P., Zucca, P., Bloomfield, D. S., McCauley, J., Gallagher, P. T., et al. \ 2013, Nature Physics, 2767
  
\bibitem[Carley et al.(2012)]{Carley2012} Carley, E.~P., McAteer, R.~T.~J., \& Gallagher, P.~T.\ 2012, \apj, 752, 36 
  
\bibitem[Cla{\ss}en \& Aurass(2002)]{Classen2002} Cla{\ss}en, H.~T., \& Aurass, H.\ 2002, \aap, 384, 1098 

\bibitem[Cho et al.(2005)]{Cho2005} Cho, K.-S., Moon, Y.-J., Dryer, M., et al.\ 2005, Journal of Geophysical Research (Space Physics), 110, 12101 

\bibitem[Cho et  al.(2007)]{Cho2007} Cho, K.-S., Lee, J., Moon, Y.-J., et al.\ 2007, \aap, 461, 1121 

\bibitem[Craig \& Brown(1976)]{Craig1976} Craig, I.~J.~D., \& Brown, J.~C.\ 1976, \aap, 49, 239 

\bibitem[Domingo et al.(1995)]{Domingo1995} Domingo, V., Fleck, B., \& Poland, A.~I.\ 1995, \solphys, 162, 1 

\bibitem[Elmore et al.(2003)]{Elmore2003} Elmore, D.~F., Burkepile, J.~T., Darnell, J.~A., Lecinski, A.~R., \& Stanger, A.~L.\ 2003, \procspie, 4843, 66 

\bibitem[Gallagher et al.(1999)]{Gallagher1999} Gallagher, P.~T., Mathioudakis, M., Keenan, F.~P., Phillips, K.~J.~H., \& Tsinganos, K.\ 1999, \apjl, 524, L133 

\bibitem[van Haarlem et al.(2013)]{VanHaarlem2013} van Haarlem, M.~P., Wise, M.~W., Gunst, A.~W., et al.\ 2013, arXiv:1305.3550 

\bibitem[Hannah \& Kontar(2012)]{Hannah2012} Hannah, I.~G., \& Kontar, E.~P.\ 2012, \aap, 539, A146 

\bibitem[Hanser \& Sellers(1996)]{Hansen1996} Hanser, F.~A., \& Sellers, F.~B.\ 1996, \procspie, 2812, 344 

\bibitem[Hayes et al.(2001)]{Hayes2001} Hayes, A.~P., Vourlidas, A., \& Howard, R.~A.\ 2001, \apj, 548, 1081 

\bibitem[van de Hulst(1947)]{Hulst1947} van de Hulst, H.~C.\ 1947, \apj, 105, 471 

\bibitem[Kerdraon \& Delouis(1997)]{Kerdraon1997} Kerdraon, A., \& Delouis, J.-M.\ 1997, Coronal Physics from Radio and Space Observations, 483, 192 

\bibitem[Klein et al.(1999)]{Klein1999} Klein, K.-L., Khan, J.~I., Vilmer, N., Delouis, J.-M., \& Aurass, H.\ 1999, \aap, 346, L53 


\bibitem[Lemen et al.(2012)]{Lemen2012} Lemen, J.~R., Title, A.~M., Akin, D.~J., et al.\ 2012, \solphys, 275, 17 

\bibitem[Mancuso \& Raymond(2004)]{Mancuso2004} Mancuso, S., \& Raymond, J.~C.\ 2004, \aap, 413, 363 

\bibitem[Mann(1992)]{Mann.I1992} Mann, I.\ 1992, \aap, 261, 329 

\bibitem[Mann et al.(1999)]{Mann1999} Mann, G., Jansen, F., MacDowall, R.~J., Kaiser, M.~L., \& Stone, R.~G.\ 1999, \aap, 348, 614 

\bibitem[Mann et al.(2003)]{Mann2003} Mann, G., Klassen, A., Aurass, H., \& Classen, H.-T.\ 2003, \aap, 400, 329 


\bibitem[Menzel(1936)]{Menzel1936} Menzel, D.~H.\ 1936, Harvard College Observatory Circular, 417, 1 

\bibitem[Minnaert(1930)]{Minnaert1930} Minnaert, M.\ 1930, \zap, 1, 209 

\bibitem[Newkirk(1961)]{Newkirk1961} Newkirk, G., Jr.\ 1961, \apj, 133, 983 

\bibitem[Parenti et al.(2000)]{Parenti2000} Parenti, S., Bromage, B.~J.~I., Poletto, G., et al.\ 2000, \aap, 363, 800 

\bibitem[Pesnell et al.(2012)]{Pesnell2012} Pesnell, W.~D., Thompson, B.~J., \& Chamberlin, P.~C.\ 2012, \solphys, 275, 3 

\bibitem[Pick et al.(2006)]{Pick2006} Pick, M., Forbes, T.~G., Mann, G., et al.\ 2006, \ssr, 123, 341 

\bibitem[Qu{\'e}merais \& Lamy(2002)]{Quemerais2002} Qu{\'e}merais, E., \& Lamy, P.\ 2002, \aap, 393, 295 

\bibitem[Ramesh(2005)]{Ramesh2005} Ramesh, R.\ 2005, Coronal and Stellar Mass Ejections, 226, 83 

\bibitem[Roberts(1959)]{Roberts1959} Roberts, J.~A.\ 1959, Australian Journal of Physics, 12, 327 

\bibitem[Robinson(1985)]{Robinson1985} Robinson, R.~D.\ 1985, \solphys, 95, 343 

\bibitem[Saito et al.(1977)]{Saito1977} Saito, K., Poland, A.~I., \& Munro, R.~H.\ 1977, \solphys, 55, 121

\bibitem[Schatten et al.(1969)]{Schatten1969} Schatten, K.~H., Wilcox, J.~M., \& Ness, N.~F.\ 1969, \solphys, 6, 442 

\bibitem[Schrijver \& De Rosa(2003)]{Schrijver2003} Schrijver, C.~J., \& De Rosa, M.~L.\ 2003, \solphys, 212, 165 

\bibitem[Uchida(1960)]{Uchida1960} Uchida, Y.\ 1960, \pasj, 12, 376 

\bibitem[Verma et al.(2013)]{Verma2013} Verma, A.~K., Fienga, A., Laskar, J., et al.\ 2013, \aap, 550, A124 

\bibitem[Vourlidas et al.(2002)]{Vourlidas2002} Vourlidas, A., Buzasi, D., Howard, R.~A., \& Esfandiari, E.\ 2002, Solar Variability: From Core to Outer Frontiers, 506, 91 

\bibitem[Vr{\v s}nak et al.(2002)]{Vrsnak2002} Vr{\v s}nak, B., Magdaleni{\'c}, J., Aurass, H., \& Mann, G.\ 2002, \aap, 396, 673 

\bibitem[Vr{\v s}nak \& Cliver(2008)]{Vrsnak2008} Vr{\v s}nak, B., \& Cliver, E.~W.\ 2008, \solphys, 253, 215 



\bibitem[Warmuth  \& Mann(2005)]{Warmuth2005} Warmuth, A., \& Mann, G.\ 2005, \aap, 435, 1123 

\bibitem[Wild \& McCready(1950)]{Wild1950} Wild, J.~P., \& McCready, L.~L.\ 1950, Australian Journal of Scientific Research A Physical Sciences, 3, 387 

\bibitem[Zucca et al.(2012)]{Zucca2012} Zucca, P., Carley, E.~P., McCauley, J., et al.\ 2012, \solphys, 94 

\end{thebibliography}
\end{document}